\documentclass[conference]{IEEEtran} 
\usepackage{booktabs}
\usepackage[draft=false]{defs_common}
\usepackage{defs_custom}
\usepackage{cite}
\usepackage{amsmath,amssymb,amsfonts}
\usepackage{algorithmic}
\usepackage{graphicx}
\usepackage{textcomp}
\usepackage{xcolor}
\def\BibTeX{{\rm B\kern-.05em{\sc i\kern-.025em b}\kern-.08em
    T\kern-.1667em\lower.7ex\hbox{E}\kern-.125emX}}

\renewcommand{\refname}{\vspace{-4pt}References}

\begin{document}

\title{Towards Game Design via Creative Machine Learning (GDCML)}

\author{\IEEEauthorblockN{Anurag Sarkar}
\IEEEauthorblockA{\textit{Northeastern University}\\
Boston, MA, USA \\
sarkar.an@northeastern.edu}
\and
\IEEEauthorblockN{Seth Cooper}
\IEEEauthorblockA{\textit{Northeastern University}\\
Boston, MA, USA \\
se.cooper@northeastern.edu}
}


\maketitle

\begin{abstract}
In recent years, machine learning (ML) systems have been increasingly applied for performing creative tasks. Such creative ML approaches have seen wide use in the domains of visual art and music for applications such as image and music generation and style transfer. However, similar creative ML techniques have not been as widely adopted in the domain of game design despite the emergence of ML-based methods for generating game content. In this paper, we argue for leveraging and repurposing such creative techniques for designing content for games, referring to these as approaches for Game Design via Creative ML (GDCML). We highlight existing systems that enable GDCML and illustrate how creative ML can inform new systems via example applications and a proposed system. 
\end{abstract}


\newcommand{\XFIGUREblending}{
\begin{figure}[b]
\centering
\includegraphics[width=1\columnwidth]{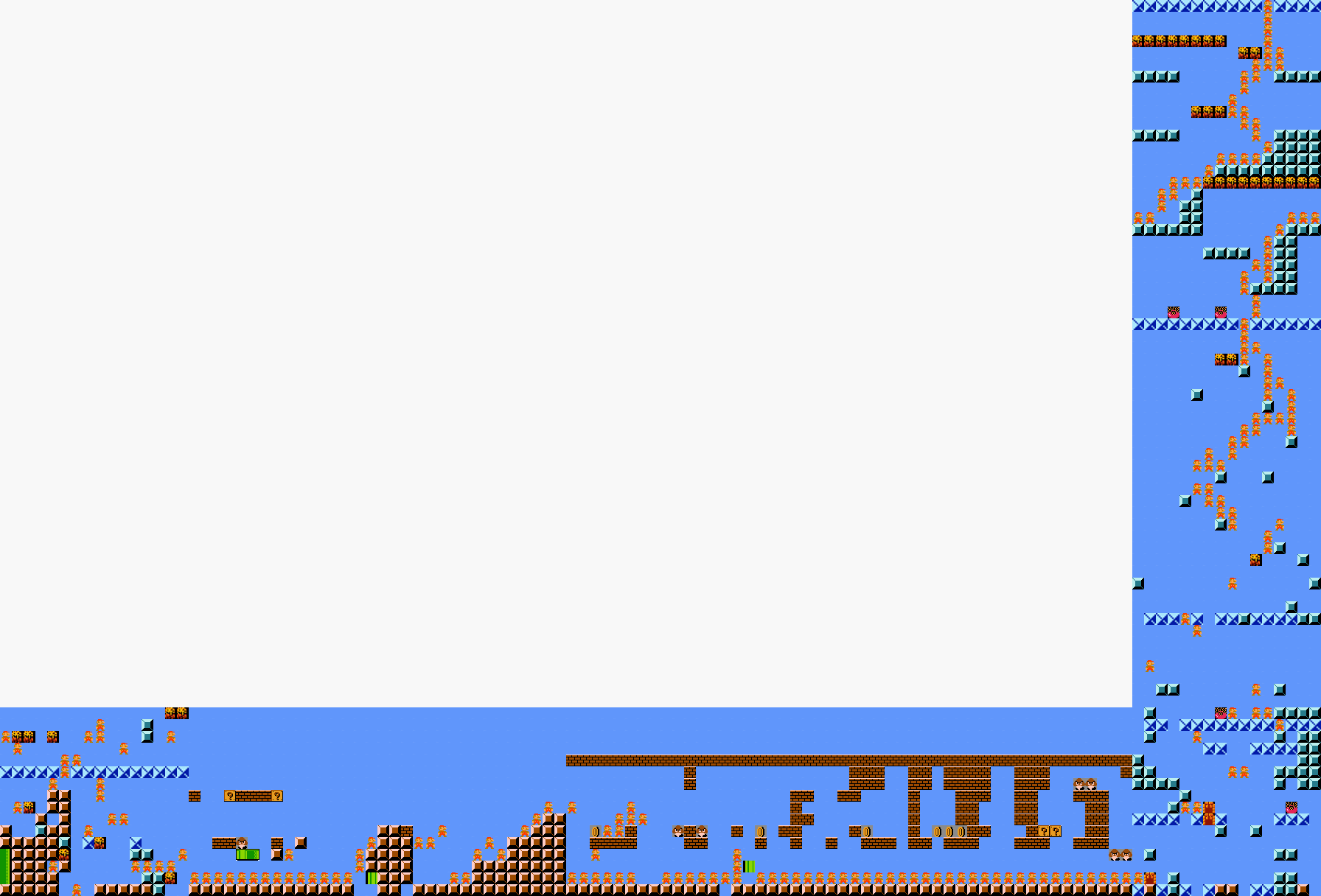}
\caption{\label{XFIGUREblending} Example level that blends \textit{Super Mario Bros.} and \textit{Kid Icarus.}}
\end{figure}
}

\newcommand{\XFIGUREsmbcont}{
\begin{figure}[t]
\centering
\begin{tabular}{c}
\includegraphics[width=1\columnwidth]{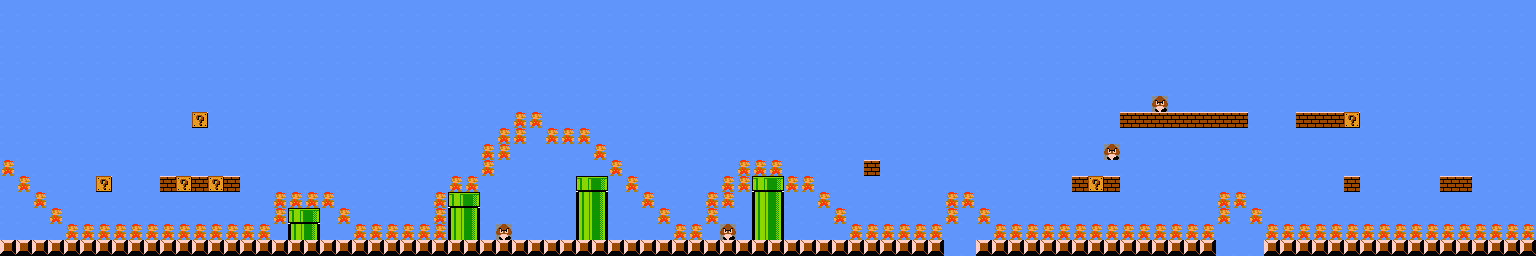}
\\
\includegraphics[width=1\columnwidth]{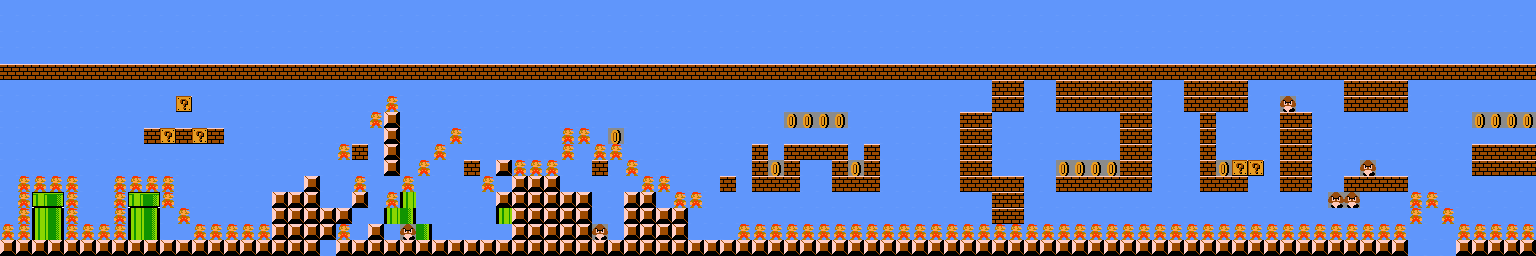}
\end{tabular}
\caption{\label{XFIGUREsmbcont} Mario levels generated from an initial given segment. The top level is generated using an initial segment taken from the original Level 1-1. The bottom level is generated using a custom initial segment.} 
\end{figure}
}

\newcommand{\XFIGUREint}{
\begin{figure}[t]
\centering
\setlength\tabcolsep{1pt}
\begin{tabular}{cccc}
\includegraphics[width=0.175\columnwidth]{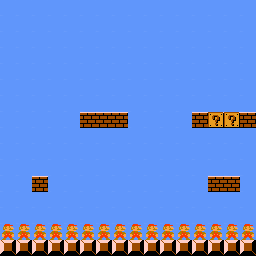}
\includegraphics[width=0.175\columnwidth]{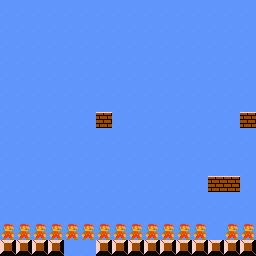}
\includegraphics[width=0.175\columnwidth]{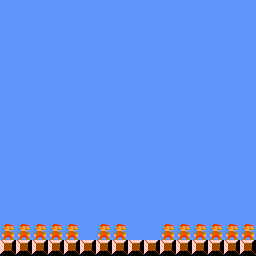}
\includegraphics[width=0.175\columnwidth]{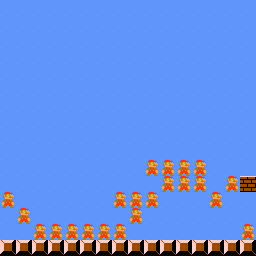}
\includegraphics[width=0.175\columnwidth]{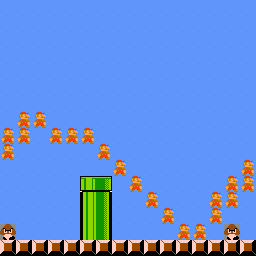}
\end{tabular}

\caption{\label{XFIGUREint} \hspace*{-0.25cm} Interpolation in Mario. Segments at each end are from the actual game.} 
\end{figure}
}

\newcommand{\XFIGUREplots}{
\begin{figure*}[t]
\setlength\tabcolsep{3pt}
\begin{tabular}{c|c|c}
\includegraphics[width=0.32\textwidth]{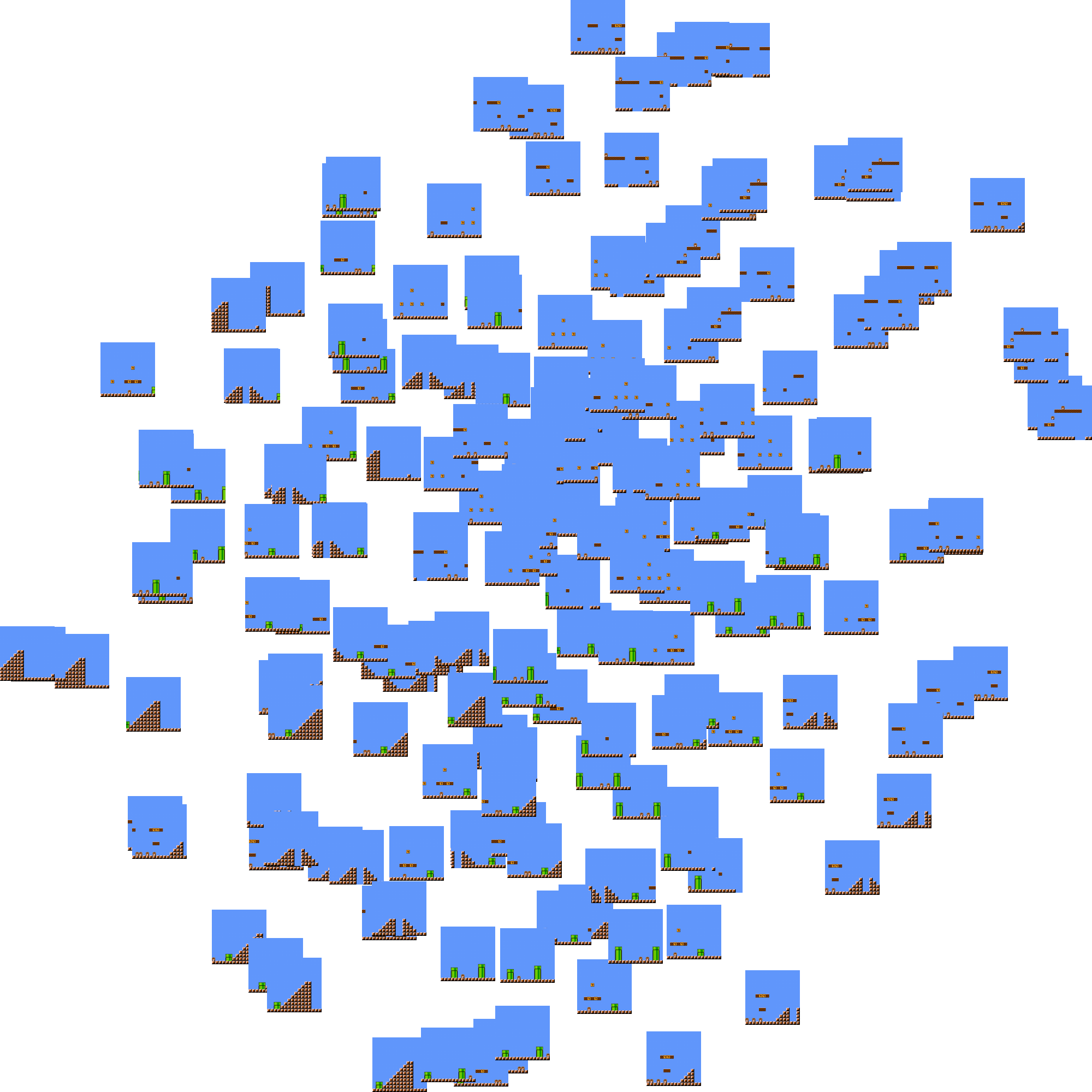} &
\includegraphics[width=0.32\textwidth]{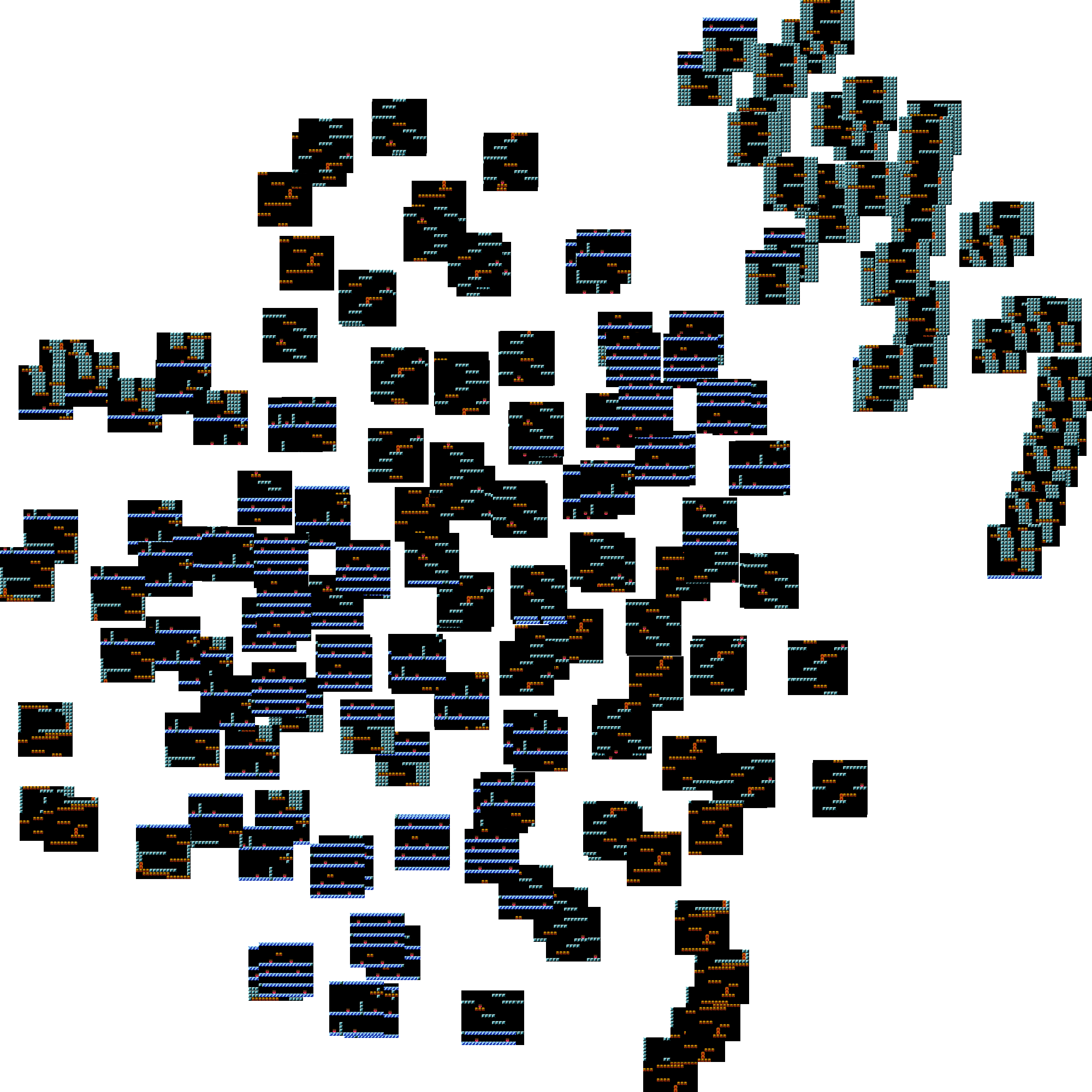} &
\includegraphics[width=0.32\textwidth]{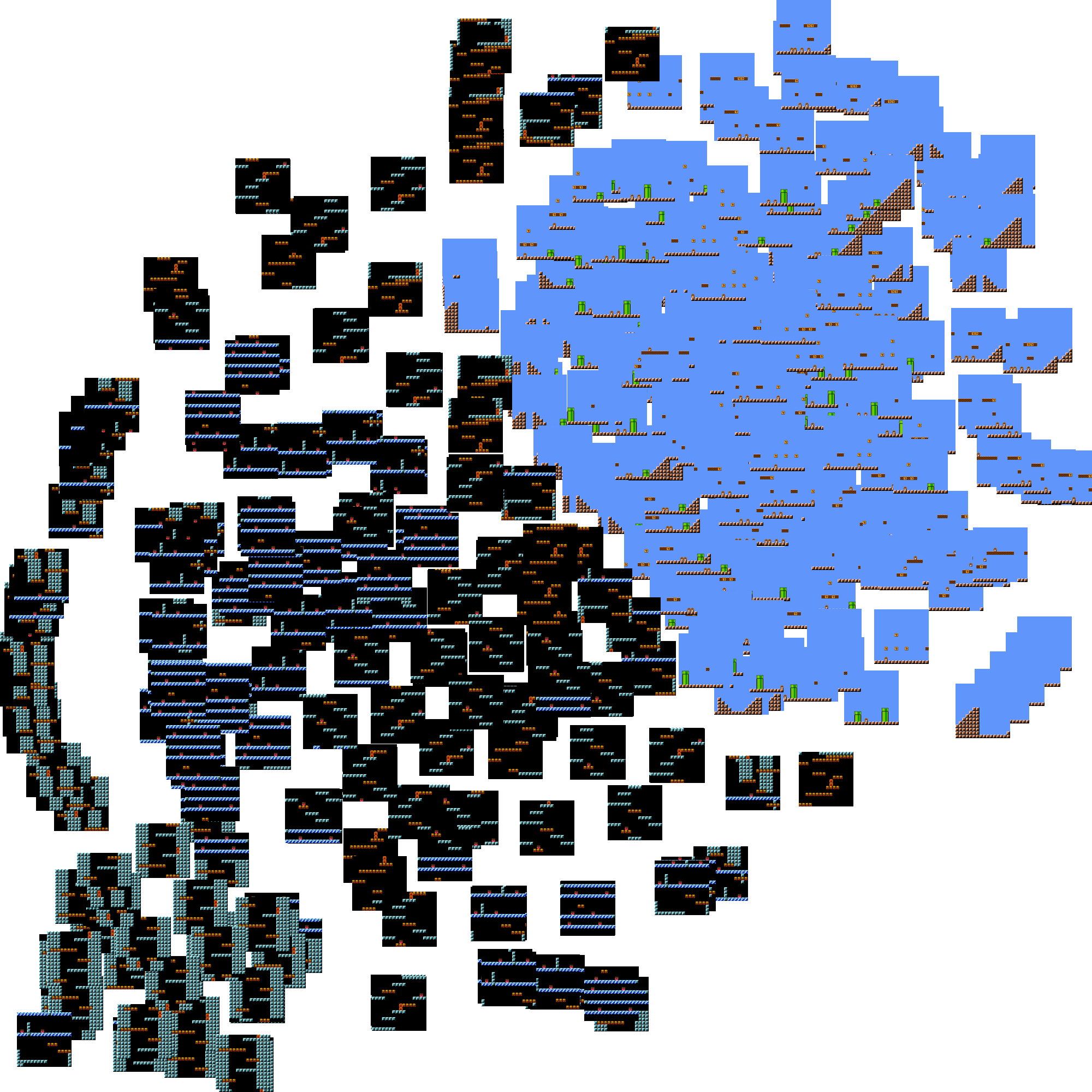} 
\end{tabular}
\caption{\label{XFIGUREplots} From left to right, t-SNE visualizations for training segments using VAEs trained on only \textit{Super Mario Bros.}, only \textit{Kid Icarus}, and both.} 
\end{figure*}
}
\section{Introduction}
Advances in machine learning, such as those in the field of computer vision and image processing \cite{krizhevsky2012imagenet, deng2009imagenet, simonyan2014very, he2016deep}, and the emergence of generative models such as GANs \cite{goodfellow_generative_2014} and VAEs \cite{kingma_auto_2013, rezende2014stochastic}, have enabled a wide variety of ML applications for performing creative and artistic tasks. In the domain of visual art, this has led to applications such as artistic style transfer \cite{gatys2016neural, huang2017arbitrary, ghiasi2017exploring}, texture synthesis \cite{gatys2015texture, ulyanov2016texture} and image manipulation and translation \cite{isola_image_2017, zhu_unpaired_2017}. Similarly, the domain of music has also been a fertile ground for such creative applications including generation using a number of different approaches \cite{briot2017deep} and in different genres and styles \cite{hadjeres2017deepbach, trieu2018jazzgan, chu2016song, zukowski2018generating}. Moreover, such \textit{creative AI} \cite{pieters_creative_2016} approaches and models have enabled the creation of generative systems and co-creative tools in both visual art \cite{Ulyanov2016fastdoodle, champandard2016semantic, bau2019gandissect, hesse2017image, opendot2016invisible, karacan2019manipulating} and music \cite{roberts2019magenta, donahue2019piano, engel2019gansynth}, helping in democratizing such AI and ML systems and better facilitating human creativity. 

However, such ML-based creative approaches have not been as widely adopted for game design. Due to the relative dearth of good training data compared to visual art and music and the added complexity of ensuring that models can produce content that is functional and playable, most generative approaches for games, often termed as procedural content generation (PCG) \cite{shaker_procedural_2016}, have primarily leveraged methods involving evolutionary search \cite{togelius_search-based_2011}, generative grammars \cite{smith_tanagra_2011}, constraint solving \cite{smith_answer_2011} and cellular automata \cite{johnson2010cellular}. More recently, a number of works have demonstrated the feasibility of using ML to build generative models for games and game content. PCG via machine learning (PCGML) \cite{summerville2018procedural} has emerged as a subfield of games research concerned with generating new game content using models trained on existing game data. A number of ML techniques have been used for this purpose such as LSTMs \cite{summerville_super_2017}, Bayes Nets \cite{guzdial2016game}, Markov models \cite{snodgrass_experiments_2014}, GANs \cite{volz_evolving_2018} and autoencoders \cite{jain_autoencoders_2016}, mostly for level generation for games such as \textit{Super Mario Bros.} \cite{supermario:nes}, \textit{The Legend of Zelda} \cite{zelda} and \textit{Doom} \cite{doom}. While these works successfully used ML to generate levels, they mostly focused on a specific game and are more comparable to simpler generative applications in visual art and music rather than the more creative applications such as style transfer and ML-based co-creativity. 

In this vein, a recent trend of more creative PCGML has emerged \cite{guzdial2018combinatorial}, focusing on applications such as domain transfer \cite{snodgrass_approach_2016, snodgrass2020multi}, level and game blending \cite{guzdial_learning_2016, sarkar_blending_2018, sarkar_controllable_2019} and generation of entire new games using ML models \cite{guzdial_automated_2018}. These new works combined with the emergence of new ML-powered game design tools \cite{guzdial2018co} signal that creative ML approaches prevalent primarily in visual art and music thus far, can be repurposed for use with ML models for game design.

In this paper, we expand on our prior exploratory work \cite{sarkar2019game} and introduce the term \emph{Game Design via Creative Machine Learning} or GDCML to refer to such techniques and models. More specifically, we use GDCML to refer to a subset of PCGML techniques that use models trained on one or more games to enable creative ML applications and affordances for automated as well as mixed-initiative design tools for game design, similar to those seen in visual art and music as highlighted previously. In the remainder of the paper, we briefly survey creative ML approaches in visual art and music and connect them to related existing and potential applications in game design, highlight existing creative ML work in games and demonstrate applications for game design via a number of illustrative examples. We also discuss the blueprint for a proposed co-creative ML-based design tool that encompasses the demonstrated applications and conclude by highlighting future directions and challenges in the field.

\section{Creative ML for Visual Art and Music}
Seminal to the eventual emergence of creative ML in visual art was the work of Gatys et al. in using convolutional neural nets to perform texture synthesis \cite{gatys2015texture} and image style transfer \cite{gatys2016neural}. This has been followed by a large body of work both in terms of research aimed at improving and building upon style transfer methods \cite{jing2019neural} as well as tools and implementations enabling users to interact and experiment with these methods \cite{Johnson2015, Ulyanov2016fastdoodle, champandard2016semantic}. Isola et al.'s pix2pix \cite{isola_image_2017} model has been particularly helpful in popularizing creative ML applications. This model learns image transformation functions between sets of image pairs and has been used in a wide variety of interactive tools and demos such as Invisible Cities \cite{opendot2016invisible}, Hesse's Image-to-Image Demo \cite{hesse2017image} as well as a number of live interfaces and installations \cite{ml4apix2pix}. Moreover, research into better understanding the inner workings of such ML models have also resulted in interesting artistic applications such as DeepDream \cite{mordvintsev2015inceptionism}, and interactive tools such as GAN Paint \cite{bau2019gandissect}. Along these lines, the software suite \textit{RunwayML} \cite{runway} enables users to work with pretrained generative models for a number of artistic tasks, demonstrating how such applications can help democratize creative ML to non-practitioners. Instrumental to the rise of creative ML in visual art has been the increase in popularity and the rapid growth in scale, complexity and expressivity of Generative Adversarial Networks (GANs) \cite{goodfellow_generative_2014} with models such as CycleGAN \cite{zhu_unpaired_2017}, SNGAN \cite{zhang2018self} and particularly BigGAN \cite{brock2018large} and StyleGAN \cite{karras2019style} being leveraged by artists such as Mario Klingemann, Helena Sarin, Robbie Barrat, Anna Ridler and Memo Akten, to name a few, to produce artworks and help usher in a new era of \textit{AI Art} \cite{hertzmann2019aesthetics}.

The domain of music has also seen much ML-based research. A wide variety of different ML approaches have been used for building generative models of music \cite{briot2017deep} using both raw audio \cite{oord2016wavenet} as well as symbolic representations \cite{yang2017midinet} and for diverse genres including Bach chorales \cite{hadjeres2017deepbach}, jazz \cite{trieu2018jazzgan}, pop \cite{chu2016song} and metal \cite{zukowski2018generating}. Like in visual art, ML research in music has also seen plenty of recent works use latent variable models such as GANs \cite{dong2018musegan, engel2019gansynth} and VAEs \cite{roberts_hierarchical_2018, engel2017latent, roberts2018learning}. These approaches all leverage the model's learned latent space to enable applications such as learning, blending and transfer of styles, instrument modeling and conditioning generation on desired attributes. Moreover, these models serve as the basis for co-creative, interactive design tools such as \textit{Magenta Studio} \cite{roberts2019magenta} and \textit{MidiMe} \cite{dinculescu2019midime} which operationalize the affordances of the ML models underneath. OpenAI's \textit{Jukebox} \cite{dhariwal2020jukebox} is a high-profile recent example of a creative ML model for music. 

Note that for both visual art and music, while the initial ML techniques enabling generative modeling were foundational, creative AI/ML did not flourish until the rise of more advanced latent variable models that enable applications such as blending, interpolation, style transfer and conditioning along with tools that operationalize and democratize them. Most current ML research in games is at the former, foundational stage. It is with a view to highlight and discuss existing and future methods to enable the latter that we write this paper.

\section{Creative AI in Game Design}
In this section, we discuss existing co-creative approaches in games. First however, we would like to draw a distinction between \textit{creative AI} and \textit{creative ML}, two terms that are often used interchangeably but we differentiate for two reasons: 1) in most uses of the term \textit{creative AI}, the underlying method more specifically uses ML and 2) to focus our scope, we wish to concentrate on co-creative game design methods and tools that use ML, separate from the various co-creative game design tools that use more general AI methods.

While the previously mentioned PCGML works are analogous to ML models for music and art in general, the closest analogs to the related tools and applications are the numerous mixed-initiative, co-creative game design tools, most of which do not employ ML. Co-creative or mixed-initiative systems \cite{yannakakis_mixed_2014} refer to those that enable human designers to collaborate with the generative system. Notable earlier examples of such systems include \textit{Tanagra} \cite{smith_tanagra_2011} for generating platformer levels, \textit{Ropossum} \cite{shaker_ropossum:_2013} for generating levels for \textit{Cut the Rope} \cite{ctr} and \textit{Sentient Sketchbook} \cite{liapis_sentient_2013} for generating strategy maps. 
More recent examples include \textit{Cicero} \cite{machado2018ai} for designing GVG-AI games, the Evolutionary Dungeon Designer \cite{alvarez2020interactive} and \textit{Baba Is Y'All} \cite{charity2020baba} for generating levels for the puzzle game \textit{Baba Is You} \cite{baba}. A related AI-based co-creative tool is \textit{Danesh} \cite{cook2016danesh} which allows users to adjust the parameters of generators and analyze their expressive range \cite{smith_understanding_2014}. Finally, though not a co-creative tool, ANGELINA \cite{cook2016angelina, cook2016angelina2} is an AI system capable of generating entirely new games using evolutionary computation. While all of these tools and systems enable the design and generation of new levels and games, they differ from the previously discussed counterparts for visual art and music in that they are not informed by ML models. That is, the systems do not leverage knowledge learned from an existing corpus of game data and subsequently are not able to harness the affordances that for example, a latent variable model would provide. In this work, we are interested in existing and potential approaches that could leverage much of the existing PCGML research to produce GDCML tools and it is this that we discuss in the next section.

\section{The Case for Creative ML for Game Design}
This paper was motivated by a recent trend of a number of works that enable a more creative form of PCGML \cite{guzdial2018combinatorial} via applications such as level and game blending, domain transfer and automated game generation. These in turn were motivated with wanting to incorporate computational creativity into ML models, specifically combinational creativity (also referred to as combinatorial creativity) \cite{boden2004creative}, the branch of creativity focused on generating new concepts, domains and artifacts from combinations of existing ones. This view of creativity contends that innovation rarely happens in a vacuum and that new ideas usually build on existing ones. Maria Popova describes this as \textit{`... the idea [is] that creativity is combinatorial, that nothing is entirely original, that everything builds on what came before, and that we create by taking existing pieces of inspiration, knowledge, skill and insight that we gather over the course of our lives and recombining them into incredible new creations'} \cite{popova2011networked}. Several other sources \cite{popova2013how, ferguson2012creativity, kleon2012steal, kowalski2019art} also highlight the prevalence of combinational creativity throughout history in both the arts and sciences. Such creativity is evident throughout the history of games as well with new games and game genres resulting from the recombination of existing ones. \textit{Metroid} \cite{metroid}, for example, combines Mario's platforming with the lock-and-key style progression of Zelda. \textit{Spelunky} \cite{spelunky} similarly melds platformer mechanics with roguelike elements. We have seen the increasing use of terms such as \textit{roguelite}, \textit{procvania} and \textit{soulslike} to describe new game genres that borrow and combine elements from multiple existing genres. Recent indie game \textit{SuperMash} \cite{supermash} lets players explicitly combine different genres to produce new games to play. Thus, imbuing ML models with combinational creativity techniques such as conceptual blending \cite{fauconnier_conceptual_1998}, amalgamation \cite{ontanon_amalgams_2010}, compositional adaptation \cite{wilke_techniques_1998} and conceptual expansion \cite{guzdial_automated_2018} could enable tools to assist in such creative forms of game design and generation. Conceptual expansion has in fact been demonstrated to be able to generate entirely new games that combine the levels and mechanics of existing games \cite{guzdial2020conceptual}.

Along these lines, Gow and Corneli \cite{gow_towards_2015} were among the first to propose a framework for blending existing games together to produce new ones. With a view towards building ML models capable of doing this, we used LSTMs \cite{sarkar_blending_2018} to blend models of Mario and \textit{Kid Icarus} \cite{kidicarus} and improved on that by using VAEs to perform controllable blending using the VAE latent space \cite{sarkar_controllable_2019}. The latter was also inspired by Volz et al.'s work \cite{volz_evolving_2018} in training GANs for generating Mario levels and using the GAN latent space to evolve variations. Since then, other works have used GANs and VAEs for PCG \cite{gutierrez2020generative, thakkar2019autoencoder, giacomello2018doom}. These approaches share similarities with analogous approaches discussed previously for visual art and music in their explicit use of the latent space and its encodings or at least their potential to do so. This in turn enables similar affordances within games (i.e. blending, interpolation, conditioned generation) as they do in other domains and thus prime these methods for serving as the foundation for co-creative GDCML tools.

To this end, Schrum et al. \cite{schrum2020interactive} recently presented a latent model-based co-creative game design tool, developing a system based on their GAN models for Mario and Zelda that allows users to design and generate levels via interactive evolution and exploration of the GAN's latent space. Thus, this tool has much in common with similar tools for visual art and music and represents the type of application we wish to highlight and build on under the GDCML definition. Another such co-creative ML-based game design tool, though not based on latent variable models but still influential to our recommendations in the following sections, is Guzdial et al.'s \textit{Morai Maker} \cite{guzdial2018co}, a Unity tool for designing Mario levels using generative models from past PCGML works \cite{guzdial_game_2017, snodgrass_experiments_2014, summerville_super_2017} as co-creative partners. In addition to these works, recent PCGML work has also looked at domain transfer methods such as Snodgrass and Onta{\~n}{\'o}n's work  \cite{snodgrass_approach_2016} in learning mappings between platformer games and Snodgrass' \cite{snodgrass2019levels} newer work that uses binary space partitioning to generate levels from a low-resolution sketch representation, an approach that seems particularly suited to inform co-creative design tools.

Overall, such tools represent promising first steps towards realizing creative ML for game design in the future, as it currently exists for visual art and music. For this to happen, existing tools need to be built upon and enhanced in terms of scope and affordances. Both \textit{Morai Maker} and Schrum et al.'s tool are restricted to a single domain. While effective, this necessarily limits the possibility space of design. Building tools that leverage existing PCGML works in blending and domain transfer described above is necessary for enabling more creative applications such as style transfer and the design and discovery of new domains and genres of games. Moreover, not much attempt has been made to borrow creative ML ideas from other domains into games. This is a missed opportunity as creative ML for visual art and music, due to its increased maturity as a field, offers many approaches and applications that can be repurposed for games. In the next section, we discuss example applications, inspired in part by those in visual art and music and in part by the existing GDCML tools above, that leverage the affordances of latent models and that we hope to implement in GDCML tools in the future.

\section{Applications}

In this section, we demonstrate example applications that we hope to implement and operationalize in future creative ML tools for game design. For some applications, we provide example figures generated from techniques and systems developed in our prior work. For these examples, we trained variational autoencoders on level data from the Video Game Level Corpus (VGLC) \cite{VGLC} for the games \textit{Super Mario Bros.} and \textit{Kid Icarus}---both classic NES-era platformers. Models were trained on 16x16 level segments using PyTorch \cite{paszke2017automatic}.  

\XFIGUREblending

\subsection{Game Blending}
Game blending refers to combining the levels and/or mechanics of two or more existing games to produce an entirely new game. Thus, it is comparable to style transfer techniques in visual art and music. Our prior work \cite{sarkar_controllable_2019} has demonstrated the feasibility of using VAEs for blending levels from separate games, motivated by implementing Gow and Corneli's VGDL game blending framework \cite{gow_towards_2015} as an ML model able to perform blending in an automated manner. The basic idea is that training a latent variable model on levels from multiple games enables it to learn a latent representation that spans and encodes levels from all games. Thus levels generated using this representation necessarily blend the properties of the original games. An example blended level is shown in Figure \ref{XFIGUREblending}. In addition to enabling users to generate such blended levels and games, tools should also allow blends to be controllable in terms of the amount of each game desired in the final blend as well as properties such as difficulty and level topology. Previous works \cite{volz_evolving_2018, sarkar_controllable_2019} have demonstrated that evolving vectors in the latent space using various objective functions can make generation controllable and find latent vectors corresponding to playable levels. Thus, GDCML tools can allow users to select different objective functions and parameters to optimize to generate desired types of blends.

\XFIGUREint

\subsection{Interpolation}
Latent variable models learn encodings of data within a continuous, latent space. When trained on game levels, such models thus enable generation of new levels that inhabit the space between existing levels via interpolation. When these levels are from different games, it enables blending as shown above but when the levels are from one specific game, we can obtain new levels not in that game. Examples of this are given in Figure \ref{XFIGUREint}. Schrum et al.'s \cite{schrum2020interactive} GAN-based tool already implements such functionality by allowing users to interpolate between two existing levels via a slider and then interact with the resulting interpolated level. We envision such features to be incorporated in GDCML tools moving forward. 

\subsection{Level Search}
This could allow designers to search for new levels given an input level and an objective. While similar to the aforementioned latent vector evolution methods, here we specifically refer to queries of the form: \textit{generate new level given input level X, metric Y and comparison condition Z.} In other words, this would enable users to generate levels that are similar/dissimilar to an input level using a given metric. Examples of this can be found in our prior work \cite{sarkar2019game}.

\XFIGUREsmbcont

\subsection{Conditioned Generation}
While above methods describe generating desired levels by searching the latent space via vector evolution, this can also be done by directly conditioning generation on either an input segment or a label, i.e. the model generates desired levels without having to use latent search. For example, a model can be trained to predict the next segment of a level given the current segment. Examples of such generations are given in Figure \ref{XFIGUREsmbcont} using an approach from our prior work \cite{sarkar2020sequential}. Such a model could be used co-creatively to generate additional content based on designer input. Alternatively, conditional VAEs allow generation to be conditioned on labels provided during training such that the model can generate levels using these labels. Such techniques could allow users to choose to generate new levels by selecting from a list of labels.


\XFIGUREplots

\subsection{Latent Space Visualization}
t-distributed Stochastic Neighbor Embedding or t-SNE \cite{maaten_visualizing_2008} is a dimensionality reduction technique that allows visualizing datapoints in a high-dimensional space by assigning them to coordinates in a lower-dimensional space, usually 2D or 3D. Within creative ML, it has been used to cluster images, paintings and audio clips based on features learned by trained models \cite{hamel_learning_2010, carr_curating_2019}. In game design however, the use of t-SNE has not been as widespread with Zhang et al.'s \cite{zhang_crawling_2018} use of t-SNE to visualize clusters of videogame moments being an example of its use in games. Similar to clustering related game moments, and analogous to clustering paintings and audio based on features, t-SNE could be used to visualize clusters of similar levels. Such visualizations, particularly when made interactive, could allow designers to explore the learned latent space and interactively search for desired levels and other content. The t-SNE visualizations of encodings of level segments from the original games are shown in Figure \ref{XFIGUREplots}. These depict clusters of level segments that share structural properties as seen by the Mario segments with pipes in the lower middle part and with pyramid like structures in the lower left part of the Mario visualization. Similar structure-based groupings can be observed for \textit{Kid Icarus} as well. Additionally, the plot for both games depicts the two games in separate clusters but more usefully, one may expect that the segments along the edge between the two clusters may be amenable for blending or combination. Thus, these visualizations can help designers interactively search the latent space for blended segments that are learned by models trained on data from multiple games. Overall, such plots help visualize the landscape of the latent space learned by the VAEs. Interactive versions of these visualizations, such as
Taylor Denouden's interactive visualization of a VAE trained on MNIST digits \cite{denouden_explorer}, could let users explore and search the latent space to find desired levels. Thus, we wish to develop such interactive visualizations as components of GDCML tools. Future work could also look into using other visualization techniques such as UMAP \cite{mcinnes_uniform_2018} which has been shown to exhibit certain benefits over t-SNE \cite{becht_evaluation_2018}.

\section{Proposed System}

We intend to implement a system that enables the above applications, focusing on 2D-side scrolling platformers from the NES era. In this section, we discuss its proposed features and functionalities in more detail. 

\subsection{Platform and Interface}
The existing tools we draw the most inspiration from are \textit{Magenta Studio} \cite{roberts2019magenta}, Google's co-creative music tool based on their MusicVAE model \cite{roberts_hierarchical_2018}; \textit{RunwayML} \cite{runway}, a recently released application that lets users run pretrained creative ML models for performing a variety of artistic tasks; and the previously described \textit{Morai Maker} \cite{guzdial2018co}. Similar to how \textit{Magenta Studio} is a suite of plugins for Ableton Live, a popular digital audio workstation, we envision our tool to consist of a suite of modules for popular game development engines, with each module corresponding to one of the applications mentioned previously. Additionally, like how \textit{RunwayML} streamlines the use of existing models for artistic applications, we intend to do the same for game design applications.

\subsection{Modules}
Our proposed system would consist of two sets of ML models---single domain, i.e. trained on a single game, and multi-domain, i.e. trained on multiple games taken together. Thus, modules pertaining to game blending would apply only for the multi-domain models while all other modules would be applicable for both. For single-domain models, users would select which game they want to work with while for multi-domain, they would select which subset of all possible games to work with. We currently envision the following modules corresponding to the previously discussed applications:

\begin{itemize}
    \item \textit{Generate} - Generate a level of user-defined length starting from a random segment. The system would pick a random vector in the latent space and forward it through the VAE's decoder until the desired size is reached.
    \item \textit{Continue} - Generate a level of user-defined length starting from a segment selected or defined by the user. Here the user supplies the segment (or selects it from a group of existing ones). This is encoded into a latent space vector and then generation proceeds as in \textit{Generate}.
    \item \textit{Interpolate} - The user defines or selects two segments, both are encoded into latent space and interpolation is done between the resulting vectors to generate new segments.
    \item \textit{Search} - This would be associated with a number of metrics and search criteria. The user defines/selects an input segment, metric and criterion which the module then uses to return a new segment based on the search results.
    \item \textit{Condition} - This would use a set of conditional VAEs each trained using separate sets of labels such as those based on level elements and level structures. In the multi-domain case, labels could correspond to separate games.
    \item \textit{Visualize} - A latent space visualizer which would allow the user to interactively explore the latent space to find segments and content to their liking.
\end{itemize}

Note that the first 3 modules above are analogous to the similarly named plugins in \textit{Magenta Studio} that respectively sample random bars from MusicVAE, extend a given music sequence using an LSTM model and interpolate between pairs of sequences using MusicVAE.

In addition to the above, for the multi-domain case, we presently plan on two blending modules:

\begin{itemize}
    \item \textit{Blend Canvas} - This is partly inspired by \textit{Beat Blender} \cite{phillips2018beat}, which allows users to interactively define and blend drum beats. We envision this module to be a similar interactive palette that lets users specify which subset of games to blend together. This would then form the single domain which can then use all of the above modules. Multi-domain models can be built either by combining single-domain models or by training a model on all available games and then extracting attribute vectors corresponding to specific games by averaging that game's latent vectors. Extracted vectors could then be combined to form the desired subset, for e.g. Vector$_{Mario}$ + Vector$_{KidIcarus}$ - Vector$_{MegaMan}$.
    \item \textit{Blend Progression} - A second blending module would allow users to select desired games to blend and then specify the blend proportions as the level progresses. For example, users could choose to blend Mario and Icarus and then specify that the first x\% of the level should be a\% Mario and b\% Icarus, the next y\% should be c\% Mario and d\% Icarus and so on. Such a module would necessitate solving interesting challenges related to generating specific blend proportions, ensuring playability, and level layout.
\end{itemize}

Tools similar to \textit{Beat Blender} such as \textit{Melody Mixer} \cite{blankensmith2018melody} and \textit{Latent Loops} \cite{loops} could also inspire modules, particularly focusing on mechanics and gameplay. This would be similar to Hoyt et al.'s \cite{hoyt2019integrating} path-based enhancements to the \textit{Morai Maker}. Compared to levels, there has been less PCGML work on mechanics with Guzdial and Riedl's \cite{guzdial_game_2017} learning of game rules from video being among the few.

\subsection{ML Models and Data}

We intend to use different variants of VAEs trained on level data from the VGLC for all our models. These would include VAEs trained to reconstruct input segments like our prior work in \cite{sarkar_controllable_2019} as well as VAEs trained to generate the next segment given the current segment like our prior work in \cite{sarkar2020sequential}. Further, we wish to explore conditional VAEs \cite{sohn2015learning} to enable some of the desired functionality described previously.

\section{Conclusion and Future Directions}
In this paper, we focused on existing and potential applications for game design via creative ML (GDCML). We discussed how creative ML approaches in visual art and music can inform similar approaches in games, highlighted applications that can be enabled by the affordances of creative PCGML models and gave a brief description of a proposed system. Apart from implementing this system, there are a number of interesting avenues to consider for future work.

\subsection{Latent Space Disentanglement}
We discussed methods of enabling controllable generation via latent vector evolution and training conditional models. Increased controllability can also be achieved by learning disentangled representations, i.e. different dimensions of the latent space encoding different attributes. Future work could thus look at leveraging such disentanglement approaches \cite{higgins2017beta}.

\subsection{Datasets}
A major challenge in PCGML is the lack of large datasets especially compared to other creative domains. Much of the advancements in these fields is due to the ability to train increasingly complex ML models on massive, publicly available datasets. While the VGLC has been a useful resource for PCGML work, it is tiny compared to traditional ML datasets. Moreover, most games only contain a handful of levels, thus necessitating models to be trained on segments of levels rather than the levels themselves to account for data scarcity. Hence, future work could look at assembling larger datasets for games and for game content not restricted to levels. A step in this direction is the Video Game Affordances Corpus \cite{bentley2019videogame}, which attempts to gather data about the affordances and interactions enabled by game levels. Sites such as the Video Game Atlas \cite{vgmaps} could also serve as a source for building game datasets.

\subsection{Blending Genres}
Most blending and domain transfer work in PCGML has focused on games of one genre, particularly platformers and even for our proposed system, we intend to focus on platformers. However, it would be interesting to blend games from different genres. What would a combination of Mario and Zelda look and play like, for example? The previously mentioned \textit{SuperMash} attempts to do this by treating one genre as primary and the other as secondary, and draws elements from the latter to incorporate into the former. Can we build systems that can help design such games? This likely requires new forms of game representation that can simultaneously encode information about different genres. For example, platformers are amenable to text-based representations while dungeons are better represented using graphs. Are there representations that offer the best of both these worlds? Investigating such questions could help bring GDCML approaches closer to their counterparts in music and visual art.

\bibliographystyle{IEEEtran}
\bibliography{refs_gdcml}

\end{document}